\documentclass[12pt]{article}

\def\thebibliography#1{\section*{References}\list
 {}{\setlength\labelwidth{1.4em}\leftmargin\labelwidth
 \setlength\parsep{0pt}\setlength\itemsep{0pt}
 \setlength{\itemindent}{-\leftmargin}
 \usecounter{enumi}}}

\begin{document}

{\bf GB2 0909+353: One of the Largest Double Radio Source}

\begin{center}
by
\end{center}

\begin{center}
M. Jamrozy \& J. Machalski
\end{center}

\begin{center}
Astronomical Observatory, Jagellonian University, ul. Orla 171,\\ 30--244 Cracow, Poland \\
e--mail: jamrozy@oa.uj.edu.pl machalsk@oa.uj.edu.pl
\end{center}

\begin{abstract}
The evidence are given that the radio source GB2 0909+353 (GB2 catalogue: Machalski 1978; ICRS 2000.0 coordinates: 09 12 51.7, +35 10 10) is likely one of the largest classical doubles known, though its optical identification is not certain. Our deep VLA observations at 5 GHz did not reveal a radio core brighter than 0.5 mJy/beam at this frequency. Thus a distance to the  source is evaluated using photometric -- redshift estimates of the faint galaxies in the optical field. The equipartition magnetic field and energy density in the source is calculated and compared with corresponding parameters of other `giant' radio sources known, showing extremely low values of both physical parameters of the source investigated. On the other hand, the age of relativistic electrons, and the advance speed of the `hot spot' in the source are typical for much smaller and brighter 3CR sources.
\end{abstract}

{\bf Key words:} radio continuum: galaxies -- galaxies: active -- galaxies: individual: GB2 0909+353

\section{Introduction}

Recently, the largest and low-brightness classical radio sources are of a 
special interest for researching a number of astrophysical problems. Especially, 
their statistical studies are of a great importance for: (1) investigations of 
the time evolution of radio sources (cf. Kaiser et al. 1997; Chy\.{z}y 1997; 
Blundell et al. 1999), (2) testing the orientation-dependent unification scheme 
(e.g. Barthel 1989; Urry \& Padovani 1995), and (3) their usefulness to probe 
the low-density intergalactic and intercluster medium (cf. Strom \& Willis 1980; 
Subrahmanyan \& Saripalli 1993; Mack et al. 1998). 

Active nucleus of the largest `giant' radio sources is likely approaching an 
endstage of its activity, therefore their studies can provide important 
informations on the properties of old AGN. The identification of a number of 
quasars among these giant sources rises a problem for the unified scheme of AGN, 
where a quasar-appearance should characterize the AGN whose jets are oriented 
closer to the observer's line of sight than those in radio galaxies. Thus, 
quasars should not have large projected linear sizes. Finally, the large angular 
size of giants allows detailed studies of their spectral evolution, rotation 
measure (RM), and the depolarization towards these sources. All the above can 
provide useful data on density of the medium at very large distances from the 
AGN.

In former decades, giant radio sources were often undetectable in radio surveys 
because they lay below the survey surface-brightness limit, even though they 
had total flux densities exceeding the survey flux limit. The new deep radio 
surveys WENSS (Rengelink et al. 1997), NVSS (Condon et al. 1998) offer a 
possibility to select unbiased samples of giant radio sources owing to their low 
surface-brightness limits. Using the NVSS survey, we have selected a sample of 
FRII-type (Fanaroff \& Riley 1974) giant candidates, which further study is in 
progress. The source GB2 0909+353 is one of them. 

In this paper we show that this source is one of the largest FRII-type sources 
with extremely low values of the equipartition magnetic field and energy 
density. The 1.4-GHz high- and low-resolution radio observations, available for 
this source, and its optical field are summarized in Sec. 2. Also our special, 
deep VLA observation 
to detect its radio core, is described there. The radio spectrum of the source 
is analysed in Sec. 3. In Sec. 4, a possible optical counterpart, the redshift 
estimate and projected size of 
the source are discussed. Finally, in Sec. 5, the equipartition magnetic field 
and energy density within the source are calculated, and compared with 
corresponding values found for other giants, as well as for much smaller 3CR 
sources.

\section{The radio 1.4 GHz map and optical field}

The source has been mapped in three recent sky surveys: FIRST (VLA B--array 1.4 GHz; Becker, White \& Helfand 1995), NVSS (VLA D--array 1.4 GHz; Condon et al. 1998), and WENSS (WSRT 325 MHz; Rengelink et al. 1997). The  low-resolution 
VLA  map at 1.4 GHz, reproduced from NVSS survey, is shown in Fig. 1 with `grey-scale' optical image from the Digitized Sky Survey (hereafter DSS) overlaid. The VLA high-resolution FIRST map confirms 
the FRII-type (Fanaroff \& Riley 1974) of the source, and gives 6.3 arc min 
angular separation of the brightest parts of its lobes. Unfortunately, no radio 
core brighter than about 1 mJy at 1.4 GHz was detected during the FIRST survey. 
Therefore, we made a follow-up, deep VLA observations to detect the
core at a higher frequency. For this purpose, the sky field centred at 
J091252.0+351000 was mapped at 4885 MHz with the B-array. That observed 
frequency and the array configuration allow to map the brightness distribution 
within a radius of about 4 arc min. With the integration time of 35 min, the 
rms brightness fluctuations were about 0.025 mJy beam$^{-1}$.Unfortunately, no radio core brighter than 0.5 mJy at 5 GHz has been detected in vicinity of the target position. 
This makes the upper limit to the core--total flux ratio of about 0.023, which 
is exactly the median ratio determined at 8 GHz for radio galaxies larger than 
1 Mpc by Ishwara-Chandra \& Saikia (1999). Thus, our observations confirm very 
low brightness of the investigated radio source at high frequencies.

\begin{figure}
\includegraphics{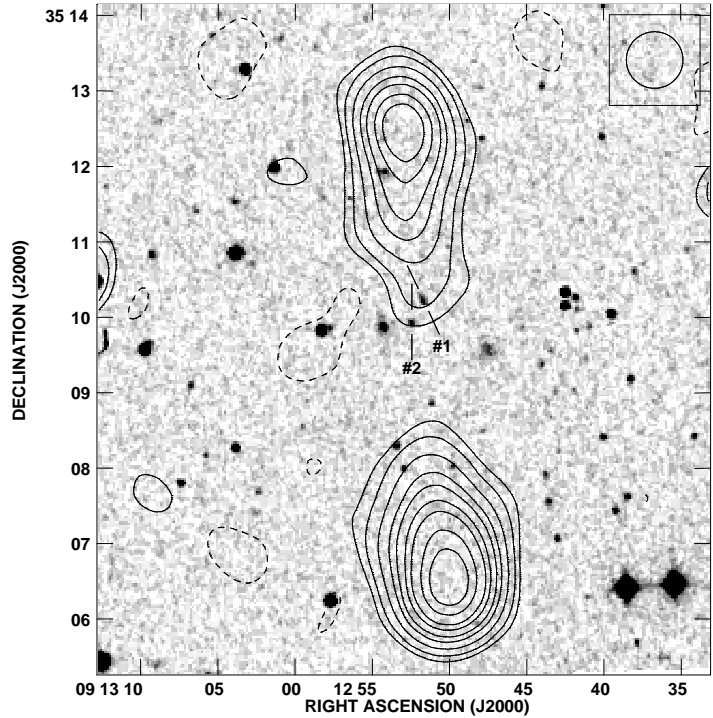}
\vspace{75mm}
\caption{The 1.4-GHz VLA map, reproduced from the NVSS survey, and overlaid on 
the optical DSS field. Contour levels are:-1,1,2,4,7,10,15,20,30,40$\times$ 
1 mJy/beam. Two possible galaxy identifications are marked}
\end{figure}

\section{The radio spectrum}

Previously the source was detected during the low-frequency sky surveys at 151 
and 408 MHz (6C2: Hales, Baldwin \& Warner 1988, and 7C: Waldram et al., 1996; 
B2.3: Colla et al., 1973), respectively, and at 1400 MHz (GB2: Machalski 1978). 
It is present, as well, in the GB6 catalogue (Gregory et al., 1996), but its 
4.85 GHz flux density may be underestimated there.
Table 1 gives the flux densities available at the frequencies from 151 MHz to 
5 GHz.

\begin{table}[htb]
\caption{Radio flux densities of the total source and its lobes}
\tabbingsep0mm
\begin{tabbing}
88888888 \=8888888899 \=11111111111 \=1010101010 \=1010101010 \= \kill
Freq.    \>Survey    \>To\'tal flux  \>So\'uth lobe \>No\'rth lobe \>Core\\
(MHz)    \>/Tel.     \>[mJy]         \>[mJy]      \>[mJy]      \>[mJy]\\
\>\\
151        \>6C2       \>8\'00$\pm$ 80 \\
151        \>7C        \>8\'20$\pm$ 60 \\
325        \>WENSS     \>5\'12$\pm$ 40 \>3\'51$\pm$ 42 \>1\'61$\pm$ 49 \\
408        \>B2.3      \>5\'11$\pm$ 72$^{a)}$ \\
1\'400     \>GB2       \>1\'28$\pm$ 28 \\
1\'400     \>NVSS      \>1\'62$\pm$ 5  \>1\'04$\pm$ 5   \>58$\pm$ 4 \\
4\'850     \>GB6       \>22$\pm$ 4 \\
4\'885     \>VLA       \>            \>             \>           \>$<$0.5
\end{tabbing}
\vspace{2mm}
Note: a) original B2.3 flux density (Colla et al., 1973) is multiplied by 
1.065, i.e. adjusted to the common flux density scale of Baars et al. (1977).
\end{table}

The radio spectrum of the total source and its lobes is shown in Fig. 2. In 
order to calculate its total radio luminosity, the spectrum is expressed by a 
functional form. The best fit of the data in column 3 of Table 1 has been 
achieved with a parabola 
$S(x)$[mJy]$=ax^{2}+bx+c$, where $x=\log\nu$[GHz]; $a=-0.493\pm 0.085, 
b=-1.097\pm 0.038, c=2.327\pm 0.030$. This fit gives the fitted 1.4-GHz total 
flux density of 143 mJy, and the fitted spectral index of $-1.24\pm 0.063$ at 
1.4 GHz. The low- and high-frequency spectral indices between fitted flux 
densities at 408 and 1400 MHz, and at 1400 and 5000 MHz, are $-0.91$ and $-1.28$, 
respectively. Such a steep slope of the spectrum at frequencies above 1 GHz 
suggests that the radio source is related to a distant galaxy. However the 
spectrum evidently flattens at low frequencies, which allows to estimate the 
lifetime of relativistic electrons in the source. Therefore, we have fitted the 
straight lines to the spectral data at low and high frequencies, estimating 
a break of the spectrum at frequency of about 760 MHz.

\begin{figure}
\includegraphics{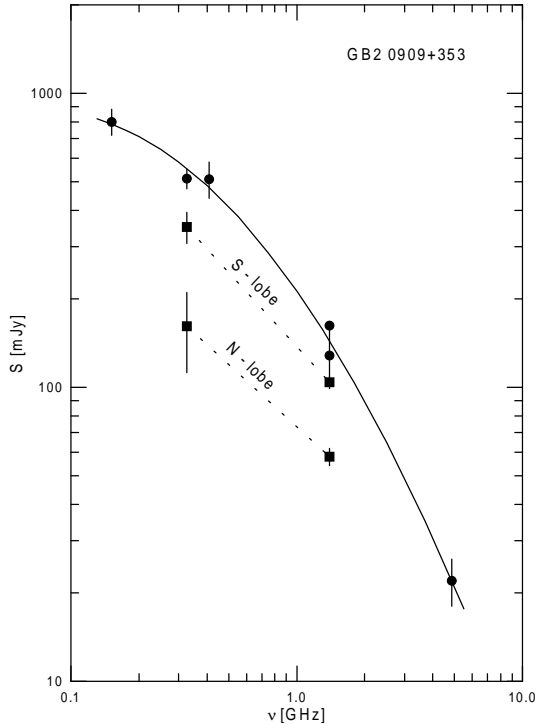}
\vspace{70mm}
\caption{Radio spectrum of the total source and its components. The solid 
curve shows the best fit of the entire spectrum}
\end{figure}

\section{The redshift estimate and size of the source}

To determine linear (projected) extent of the source, a distance to the source 
must be known. The optical field suggests that the radio source may be 
associated with one of two objects, marked \#1 and \#2 in Fig. 1. Object \#1 is 
classified as a galaxy with R=19.35 mag in the DSS. Object \#2 is not classified there because it is not visible on the Palomar Observatory Sky Survey's blue 
plates, however it is likely another red galaxy with R=19.5$\pm$0.4 mag.
Because no radio core was detected, we can only assume that any 
counterpart optical galaxy is not brighter than R=19.35 mag. Taking into 
account the Hubble relation R(z), well established for 3CR, and other radio 
galaxies (e.g. Kristian, Sandage \& Westphal 1978; Machalski 1988), the galaxy 
should be at redshift $z^{>}_{\sim}0.4$. This estimate is supported by the 
$R$-band Hubble diagram for the giant radio galaxies, recently published by 
Schoenmakers et al. (1998). On the basis of available spactroscopy of giants 
from the Westerbork sample and 7C sample of Cotter et al. (1996), they found 

\[\log z=0.125 R-2.79\]

\noindent
which, for R=19.35 mag, gives z=0.42.
This redshift estimate corresponds to the source's 
distance of about 2.6 Gpc (assuming $H_{o}=50, \Omega=1$), and linear size of 
2.43 Mpc. If so, the source GB2 0909+353 is one of about the ten FRII-type 
radio sources extended over 2 Mpc. The
radio source, investigated here, will be still over 1 Mpc (a conventional 
lower limit of size for `giant' sources) at a redshift as low as 0.11.

\section{The equipartition magnetic field and energy density}

The source is characterized by exceptionally low minimum energy density of 
relativistic particles and equipartition magnetic field. Recently, Ishwara-
Chandra \& Saikia (1999) have published very interesting statistics of the 
above parameters calculated for known radio sources extended more than 1 Mpc,
and compared them with corresponding statistics of smaller 3CR sources. 
Following Ishwara-Chandra \& Saikia, we calculate the minimum energy density 
$u_{min}$, and equipartition magnetic field $B_{eq}$ in GB2 0909+353 with the 
standard method (e.g. Miley 1980), assuming a cylindrical geometry of the source, 
a filling factor of unity, and equal energy distribution between relativistic 
electrons and protons. Integrating luminosity of the source between 10 MHz and 
10 GHz, we found $B_{eq}=0.084^{+0.040}_{-0.024}$ nT and 
$u_{min}=(0.65^{+0.55}_{-0.33})\times 10^{-13}$ erg\,cm$^{-3}$. These 
values are weakly dependent of unknown distance $D(z)$ to the source; 
$B_{eq}\propto D^{-2/7}$, $u_{min}\propto D^{-4/7}$. This means that varying 
redshift by 2 (100 per cent), one can expect a change of $B_{eq}$ by 18 per 
cent, and $u_{min}$ by 33 per cent, only.

$B_{eq}$ and $u_{min}$ values, found for GB2 0909+353, are rather extremal 
among the corresponding values for known giant sources. Ishwara-Chandra \& 
Saikia have showed that $B_{eq}$  in almost all radio sources larger than 1 Mpc 
is less than equivalent magnetic field of the microwave background radiation, 
i.e. $B_{eq}<B_{iC}\equiv0.324(1+z)^{2}$[nT]. They also have showed that 
oppositely, 
$B_{eq}>B_{iC}$ for more luminous and smaller 3CR radio sources, suggesting that  the inverse-Compton losses dominate the synchrotron radiative losses in the 
evolution of the lobes of giant sources.
Our calculation shows that the $B_{iC}/B_{eq}$ ratio for GB2 0909+353 may vary 
from 7.6$\pm$2.9 (if z=0.4) to 4.6$\pm$1.7 (if z=0.11), respectively, and 
$B^{2}_{eq}/(B^{2}_{iC}+B^{2}_{eq})$, which represents the ratio of the 
energy losses by synchrotron radiation to total energy losses due to both the 
processes, may vary from 0.017$\pm$0.013 (if z=0.4) to 0.044$\pm$0.035 
(if z=0.11), respectively. In Figs. 3(a) and 3(b) we plot these values on the 
diagrams reproduced from the paper of Ishwara-Chandra \& Saikia, and showing 
the above ratios as a function of the linear size of radio source. 
Similarly, the value of $u_{min}$ for GB2 0909+353 vs. linear size, and 
corresponding values for other giants from that paper, are plotted in Fig. 3(c). The solid line marks the expected relation $\log u_{min}=-(4/7)\log D+const$.
The loci of the source GB2 0909+353 on these diagrams fully support our thesis 
that this source is one of the largest radio sources with extremely low energy 
density and equivalent magnetic field.

\begin{figure}
\includegraphics{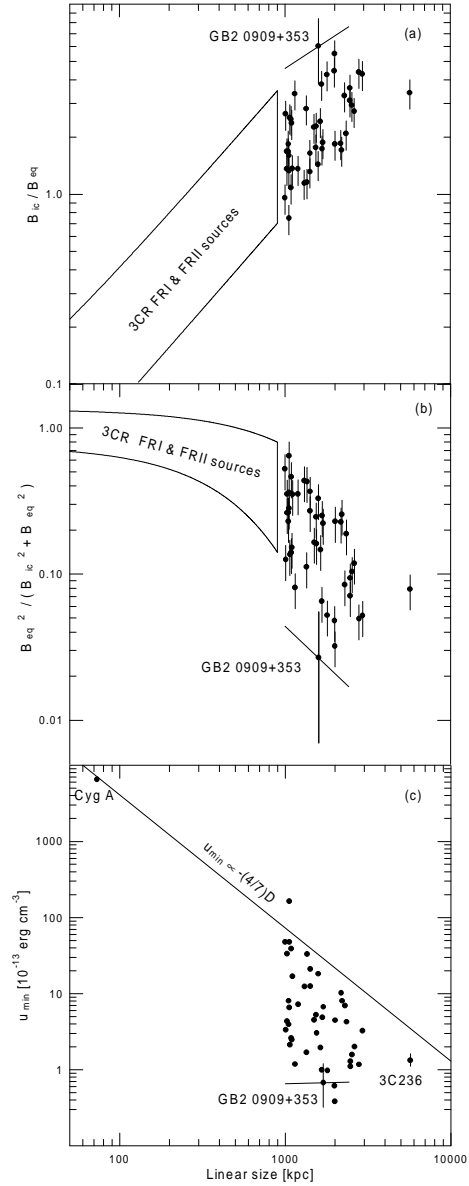}
\vspace{130mm}
\caption{(a) The ratio $B_{iC}/B_{eq}$, (b) the ratio 
$B^{2}_{eq}/(B^{2}_{eq}+B^{2}_{iC})$, and (c) $u_{min}$ plotted against the 
projected linear size of `giant' sources from the sample of Ishwara-Chandra 
\& Saikia (1999). 3CR double sources smaller than 1 Mpc, used by them, lay in 
the area confined by the solid lines in panels (a) and (b). The locus of the 
source 0909+353 is marked in each panel. The diagonal bars indicate its possible linear-size range because of unknown redshift, while vertical bars indicate the 
rms error of particular ordinate value. The solid line in panel (c) indicates 
the dependence expected in standard synchrotron source. $u_{min}$ value of 
Cyg A is taken from Alexander, Brown \& Scott (1984).}
\end{figure}

The age of relativistic electrons, derived from the value of $B_{eq}$ and $B_{iC}$ at the 
break frequency of 760 MHz may vary from $3.4\,10^{7}$ yr (if z=0.4) to 
$9.8\,10^{7}$ yr (if z=0.11), respectively. Assuming that (1)  a redshift of the 
source is between the above values, and (2) the main axis of the source is close to 
the plane of the sky, and thus any physical distance from the center should not 
differ significantly from the projected one (this assumption 
is based on the high symmetry of the source) -- a distance from the central 
galaxy to the brightest spots in the radio lobes  should be within 640 kpc and 
1540 kpc, the break frequency, characterizing the total radio spectrum, can be 
related to any distance between the above values, and the mean hotspot 
separation speed (resulting from this distance and the age of relativistic 
electrons) should be within 0.02$c$ and 0.15$c$. The range of these values is 
essentially the same as those found for much smaller, double 
3CR radio sources (cf. Alexander \& Leahy 1987; Liu, Pooley \& Riley 1992), 
however the lower value seems to be much more likely in view of the strong 
positive correlation between the separation speed and 178-MHz luminosity, found 
by these authors.  Detailed spectral observations and the ageing analysis of a 
sample of giant double sources are necessary to check whether this correlation 
holds for the largest radio sources.

Concluding, we argue that the source GB2 0909+353 is one of the largest, 
low-brightness, and distant classical double radio source. 

\vspace{5mm}
{\bf Aknowledgements.}
The VLA is operated by the National Radio Astronomy Observatory (NRAO) for Associated Universities Inc. under a licence from the National Science Foundation of the USA. We acknowledge usage of the Digitized Sky Survey which was produced at the Space Telescope Science Institute based on photografic data obtained using the Oschin Schmidt Telescope on Palomar Mountain and the UK Schmidt Telescope. We thank the anonymous referee for the valuable comments.

\end{document}